\documentstyle[sprocl,epsfig,psfig]{article}

\def\be{\begin{equation}}
\def\ee{\end{equation}}
\def\beq{\begin{equation}}
\def\eeq{\end{equation}}
\def\bea{\begin{eqnarray}}
\def\eea{\end{eqnarray}}
\def\barr{\begin{array}}
\def\earr{\end{array}}

\begin{document}
\title{Synergy between the LHC and the ILC}
 \author{ROHINI M GODBOLE}
 \address{$^1$Center for High Energy Physics, IISc, Bangalore 560012, India\\}


\begin{flushright}
{IISc-CHEP/17/04}\\
{hep-ph/0501256}
\end{flushright}
\begin{center}
\vspace{0.5cm}

{\Large \bf
Synergy between the LHC and the ILC~\footnote{Plenary talk presented at the
International Conference on Linear Colliders, April 18-23, 2004, Paris.}
}\\[2.00 cm]

{\large R. M. Godbole}
\\[0.5 cm]
{\it
Centre for High Energy Physics, Indian Institute of Science\\
Bangalore, 560012,India\\[2cm]}
\end{center}

\begin{center}
{\large ABSTRACT}

\vspace{0.5cm}
\end{center}

{\noindent\normalsize
In this talk I explain in brief the motivations behind undertaking
a study of the LHC-ILC interplay. I will give information about
the activities of the LHC-LC study group as well as the study group document.
I will illustrate the scope of the document by taking a few examples from
the document which has appeared on the archives~\cite{Group:2004hn} since
this talk was presented.
}
\newpage

\maketitle\abstracts{
In this talk I explain in brief the motivations behind undertaking 
a study of the LHC-ILC interplay. I will give information about
the activities of the LHC-LC study group as well as the study group document.
I will illustrate the scope of the document by taking a few examples from 
the document which has appeared on the archives~\cite{Group:2004hn} since
this talk was presented.
}
\section*{Introduction}

The field of high energy physics is poised at a very interesting 
juncture at present, with the Large Hadron Collider (LHC) ready to start 
colliding protons on protons at  CERN in the year 2007 and the two detectors
CMS and ATLAS getting geared to start studies at the LHC. The  particle physics 
community pins its hopes on the LHC to shed light on the crucial question of 
the origin of mass of {\it all} the fundamental particles. At the same time, 
there is 
now a world wide consensus that the next big facility in High Energy Physics 
should be an International $e^+e^-$ collider which has to be a Linear Collider;
the  International Linear Collider: the ILC. A look at the history of high 
energy colliders brings up many examples of the complementary roles played by 
the hadronic and leptonic colliders in furthering the frontiers of our 
knowledge through an interplay and feedback between the two types of 
colliders. There are many examples in the past where a new particle has 
been discovered at one machine, and its properties have been studied in 
detail with measurements at the other. Similarly, experimental results 
obtained at one machine have often given rise to predictions that have led 
to new searches at the other machine, resulting in ground-breaking discoveries.
The current state of play in the field of High Energy Physics, the long time 
that has to necessarily elapse between the  conceptual design of an 
accelerator and the actual commissioning of experiments and 
the very high stakes in physics studies at the next generation
colliders; both on the economic and physics front, make it imperative that
we as a community assess the desired energy, the luminosity and 
{\bf the timing} of this planned ILC vis-a-vis the Physics Goals that we all 
are hopeful to reach at the LHC.  

For a critical assessment of the above issues, a very close interaction between 
the experimental communities involved in the LHC and the ILC studies is 
absolutely essential. A LHC/LC study group was formed in 2002 in the ECFA/DESY
framework with the aim of achieving this. Given the 
worldwide nature of the LC study groups~\cite{wwidestudy} and the International
nature of the LHC itself, such a study group also took a worldwide character 
very soon. At present the LHC/LC study group contains about 116 members 
which includes theorists, members of the CMS and ATLAS collaborations, members
of all the LC study groups~\cite{tesnjlc} as well as contact persons
from the Tevatron. The study group had series of meetings over a period
of two years, including a very serious and vigorous activity in the 
TeV Collider Workshop at Les Houches in 2003. More information on the
activities of the Working Group can be obtained from the 
webpage: www.ippp.dur.ac.uk/$\sim$georg/lhclc. The International Linear 
Collider Steering Committee (ILCSC), also unanimously supported the idea
of such a study group.

Since the Physics case for  
the LHC and the ILC, each with its own virtues, has been clearly established,
this working group basically wanted to look in detail how the two can 
complement each other. The aim was to study how information obtained at both
the machines can be used most optimally to get more conclusive and effective
answers to the very fundamental questions of the structure of the space and 
time that the HEP community is asking at present. The aim was not to compare 
which of the two colliders can do {\it better}, but rather how  
measurements at the ILC might give pointers to new bench marks for measurements
to be performed at the LHC. While the information from the ILC may not affect
the triggering it may certainly affect the luminosity/detector upgrades as well
provide a yet sharper focus to the LHC studies by eliminating some of the
possibilities of extensions beyond the SM and/or narrowing down the 
allowed parameter space in the context of a specific model. Indeed, the 
upgrades at the Tevatron have benefited from information obtained from the 
precision measurements at the LEP. 

The study group therefore aimed at identifying issues where the cross-talk
between the two can increase the utility of {\it both}. For the 
sake of definiteness it was assumed that the LHC will run for about 20 years 
and that the ILC will come into operation after the LHC has been running for 
a few years. This will be possible if the somewhat aggressive time table for 
the ILC, envisaged by the  ILCSC,
can be adhered to. The possibilities of the cross-talk were analysed assuming
a generic situation that the Tevatron and the LHC will see new physics, but 
the nature of new physics will not be entirely clear. 

At the time of the conference the LHC/ILC study group document was close to its
final form and is now available as a hep-ph preprint~\cite{Group:2004hn},
submitted for publication. The document contains work done by about 116 
authors, discussed over seven meetings, contains about 470 pages. 
A large number of examples of complementarity and cross-talk  between the
two colliders have been identified and studied. The studies show that 
indeed there are many scenarios where the LHC experiments can benefit from 
knowledge obtained from the ILC and vice versa. While no examples were found 
where the triggering at the LHC could be affected by the input from studies at 
the ILC, points for further studies were identified which may reveal 
such examples.

\section*{LHC/ILC interplay}
We begin the discussion of the interplay by reminding ourselves of the
different virtues of the two machines. The strongest point about the LHC, 
which is a $pp$ collider with $\sqrt{s} = 14$ TeV is of course that it is 
already under construction and has a  large mass reach for direct discoveries 
of new physics. Even though the initial state kinematics of the constituent
collisions in the hard interaction is not known, conservation of the 
transverse momentum $P_t$ allows 
to make good kinematical measurements. The composite and the strongly
interacting nature of the colliding protons implies that at the LHC one will 
always have underlying events and the QCD backgrounds need to be known 
accurately.  The ILC, which envisages $e^+e^-$ collisions with a cm energy
$\sqrt{s} = 0.5$--$1$ TeV, certainly will have a lesser reach in energy but has
the strong point of doing high precision measurements due to the cleaner
environment and precise knowledge of the kinematics and polarisation
of the initial state. Backgrounds are of course much less severe, the 
options of $\gamma \gamma, \gamma p$ collisions open up new avenues to study 
the physics of the EW symmetry breaking and the physics beyond the SM. The 
high precision of measurements possible at the ILC can make it sensitive 
to the {\it indirect effects} of the same particles which the LHC expects to be 
able to produce {\it directly}. Thus information from a lower energy ILC 
can feed back into studies at the LHC. This is indeed the simplest form of 
synergy. Indeed we have seen the interplay between the top quark mass 
estimation from the precision EW measurements and direct measurements  
from the Tevatron. We also see the impact of $m_t$ measurement from the 
Tevatron on the limits for the SM Higgs boson. Precision measurements from 
the ILC can thus tell sometimes LHC where to focus the effort. Precision 
measurements at the LHC, though not impossible, are difficult at the LHC and 
hence will be possible only after 
a few years. These can thus benefit (and help us realise the LHC potential
completely) from a feedback from the ILC. The capabilities of the ILC are of
course clearly not restricted only to precision measurements, but also
include making discoveries which at times will be difficult or impossible 
at the LHC. Qualitative statements made above are obvious and have to be 
supported by quantitative studies, which are indeed present in the 
document.

Specifically three different kinds of scenarios for the cross-talks 
have been discussed in the document. First is the simple 'linear' addition 
of the utility of both the machines, where the ILC data can help clear up the 
underlying structure of the new physics of which the Tevatron and LHC will 
offer us a glimpse. This scenario does not require, for example, an overlap
in the operational period of the two accelerators. Second scenario involves a
higher level of synergy, where a combined analysis/interpretation of the LHC 
and the ILC data can make the total bigger than the sum and help, 
in particular, reduce the model dependencies in the analysis. This is not 
unlike the effect of  a reanalysis of the older JADE data  on the 
determination of $\alpha_s$. While this may not require a strict overlap of 
the two machines, given the time it takes to develop the data analysis tools 
and the huge amount of the LHC data that would need to be archived, least
amount of time difference, including a negative one, in the operational
lives of the two will help matters. If there is indeed time overlap, data from 
the ILC could influence the second phase of the LHC by providing input to the
upgrade options for the LHC machines and detectors. Again, not wholly unlike 
the synergy between the LEP and Tevatron, where the upgrade of Tevatron 
detectors has been affected by the data from LEP. The benefits from both 
the machines will be maximal in the last case. I will take different examples 
from the document~\cite{Group:2004hn} in the subjects of EW symmetry breaking; 
establishing and understanding the Higgs mechanism, Supersymmetry(SUSY) 
including the Supersymmetric Higgses and last but not the least 
the issues in EW symmetry breaking which are alternates to the 
Higgs mechanism such as dynamical symmetry breaking or alternatives to 
SUSY such as Little Higgs etc. It goes without saying that what is 
presented here is a very {\it small and incomplete} sample of the results in
the document.

\subsection*{EW Symmetry breaking: the Higgs mechanism}
All of us are quite sure that if the SM Higgs exists,
the LHC will be able\cite{atlastdr} to observe the SM Higgs
and afford measurements of its various properties such as the width, relative 
couplings etc., to an accuracy of about $10$--$15 \%$ by the end of the high 
luminosity run. The ILC~\cite{tesnjlc} will of course be capable of profiling
the SM Higgs with a great degree of accuracy even in the low energy, moderate
luminosity option, {\bf except for the $t \bar t H$ coupling {\it and} a full
reconstruction of the Higgs potential.} One of the questions addressed in  the 
document~\cite{htt} in this context  was to see if and how this situation can
be improved with a LHC/ILC cross-talk. 

\noindent \underline{Top Yukawa coupling measurements}

\vspace{0.1cm}

\noindent A good measurement of the $H t \bar t$ coupling,  
$g_{t \bar t H}$, is quite essential to be able to confirm the Higgs 
mechanism as $\it the$ origin of fermion masses.  It  will be  
accessible at both the LHC and the ILC, through a study of $t \bar t H$ 
production. While couplings of the Higgs to all the other fermions can be 
measured to a high precision at a low energy ($< 500$ GeV) ILC, a precision 
measurement of  $g_{t \bar t H}$  will require 
$\sqrt{s} \simeq 800$ -- $1000$ GeV. The LHC measurement is 
{\bf model dependent}.  Combining  the ILC
precision measurements of the branching ratios of the Higgs into different 
channels, along with LHC measurements of the 
$\sigma (pp \rightarrow H + X) \times {\rm B.R.}$ for various final states,
one can determine the $t \bar t H$ coupling in a model independent way.
The dashed line in Fig.1 shows  the relative error  of measurement 
of $g_{t \bar t H}$  that  can be achieved with 
the LHC and a low energy, moderate luminosity  ILC. The lower solid curve
shows that for an ILC to achieve a similar accuracy on its own will require a 
much higher luminosity and of course a higher energy which may be possible in
the second stage of an ILC. So this is an example 
where the cross-talk increases utility of both, the LHC and a low energy ILC.
\begin{figure}[!t]
\centering{\psfig{file=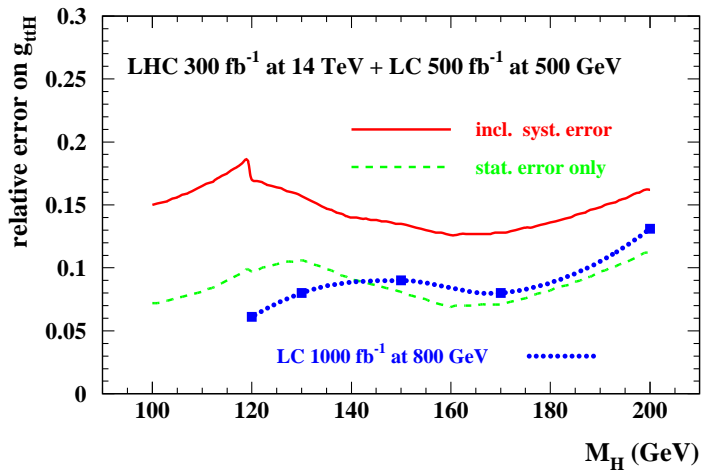,height=2.5in,angle=0}}
\caption{\label{tth} Relative error on measurement of $g_{t  \bar t  H}$ 
from LHC + ILC.}
\end{figure}

\noindent \underline{Higgs self coupling measurements}

\vspace{0.1cm}

\noindent 
The complementarity of the two machines for these two measurements is truly
remarkable. For low Higgs masses $m_H < 140$ GeV, a $500$ GeV ILC offers 
the best chance with a luminosity of 1 ab$^{-1}$ with LHC offering no 
possibilities at all. On the other hand for a heavier Higgs the situation 
is reversed. However, even then the reconstruction of the scalar potential
will require a luminosity upgrade of the LHC and precision information on 
the $g_{HWW}, \Gamma_H$ and $g_{t \bar t H}$ from a low energy ILC will be 
an important input. The present study~\cite{bauer} explores the possibilities 
of the LHC/ILC synergy here.  This is an example where the LHC/ILC studies 
have identified further work that needs be done to address issues of 
systematic uncertainties etc. so as to confirm the conclusions arrived here.. 

\subsection*{Strong EW symmetry breaking}
If a higgs boson with a mass less than the upper bounds set by the unitarity
arguments is not seen at the LHC, it would imply that the EW symmetry breaking
dynamics has to be tested in $W/Z$ scattering processes. Such studies are also
interesting in the context of the new ideas such as the Little Higgs models.
Separate studies exist for the LHC and the ILC which estimate how well the $WW$ scattering probes this dynamics. Due to the different theoretical and 
experimental approximations it is not possible to combine the results of these
analyses numerically. One can make the following qualitative statement. The 
LHC and the ILC are sensitive to different and/or complementary channels.
The LHC clearly has the sensitivity to higher mass resonances, but the ILC has 
the ability, for example, to separate the $WW$ final state from the $ZZ$ final 
states. In general, both at the LHC and the ILC, the studies find large
correlations among  different  parameters of the model of  the strong EW 
symmetry breaking. Combined analysis of the LHC and the ILC data, would 
clearly reduce these correlations. A full and efficient use of the LHC data 
would {\it require} detailed information on multi-fermion final states such as
the angular distributions etc. from the ILC. Combination with data from 
a  {\it sub TeV} ILC will be absolutely essential in disentangling the states 
that the LHC will be capable of producing. Combined LHC/ILC studies are now in 
progress~\cite{Barklow}.

\subsection*{Supersymmetry}
Supersymmetry is arguably the 'standard' 'BSM: Beyond the Standard Model' 
physics. If TeV scale Supersymmetry exists, signal for some Supersymmetric 
particle is sure to show up  at the LHC. The LHC studies have already shifted 
gear from exploring the 'discovery' potential to exploring the 
SUSY 'spectroscopy'~\cite{atlastdr}.  The latter, which consists of determining
the properties  of SUSY particles such as their masses and couplings, has been
performed mainly in a model dependent manner. Given the plethora of SUSY models
which correspond to different SUSY breaking mechanisms and essentially reflects
our ignorance of the SUSY breaking mechanism~\cite{ourbook}, model independent
analyses will be welcome in the context.

\vspace{.3cm}

\noindent \underline{Determination of sparticle masses, SUSY parameters: LHC/ILC synergy}

\vspace{0.05cm}

\noindent 
At the LHC, the SUSY signal will be caused by the sparticle production in
pairs and the decay of these involving long decay chains ending, for the 
R-parity conserving case, in the lightest Supersymmetric particle, 
the LSP $\tilde\chi_1^0$. Since the LSP is 'lost' the sparticle mass 
determination at the LHC will be basically done by using the 'edges' from the 
decay chains, eg.  $\tilde \chi_2^0 \rightarrow \tilde \chi_1^0 l^+ l^-$.  This procedure gives rise to an obvious problem of rather strong correlation between
the determined sparticle mass and  that of the LSP. Thus the accuracy of the
sparticle mass determination will be strongly affected by the precision with 
which the LSP mass is known. The analyses clearly show that the error in the
gluino mass $\Delta m_{g} \sim \Delta m_{\tilde \chi_1^0}$, the exact relation
depending upon the experimental analysis such as the jet scale uncertainty.
The issue of how the mass measurements at the LHC could be improved by 
information available from the ILC, was studied~\cite{spartmasses} using 
the results of the ATLAS analysis of the completely simulated events 
for the point SPS1a~\cite{SPS}. This point  has a light sparticle spectrum 
and both the LHC and the ILC have reach for a number of lighter sparticles.

\begin{minipage}{5cm}
\includegraphics*[scale=.25]{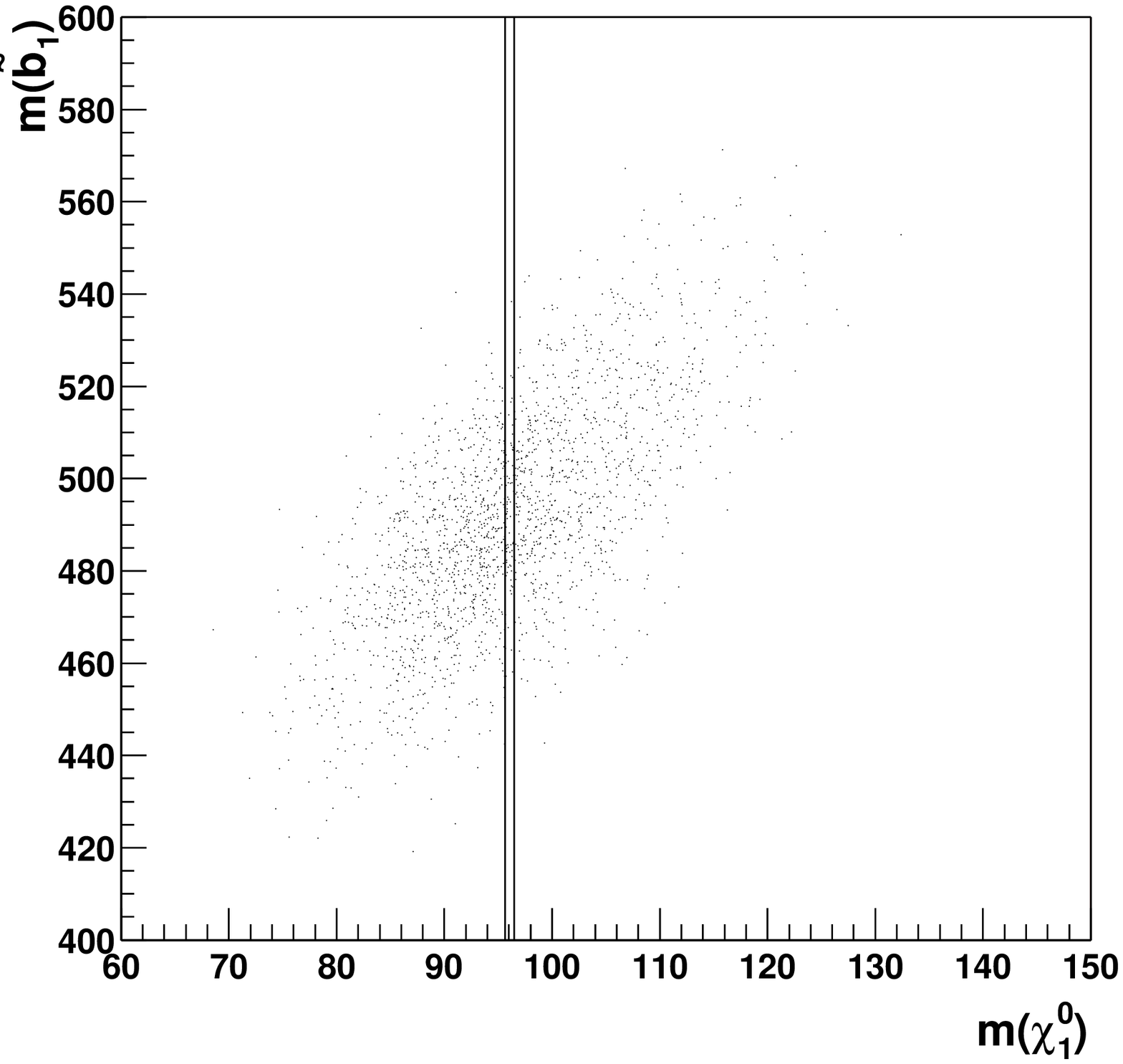}
{\footnotesize Figure 2 : Mass correlation plots. Dots: LHC alone.
Vertical bands: Fixing $m_{\tilde{\chi}_1^0}$ to within
$\pm2\sigma$ with LC input ($\sigma=0.2\%$).}
\end{minipage}
\hspace{1.0cm}
\begin{minipage}{5.0cm}
\begin{tabular}{cccc}
\hline\hline
 & LHC & LHC+LC  \\
\hline
$\Delta m_{\tilde{\chi}_1^0}$& 4.8 & 0.05 (LC input) \\
$\Delta m_{\tilde{\chi}_2^0}$& 4.7 & 0.08 \\
$\Delta m_{\tilde{\chi}_4^0}$ & 5.1 & 2.23 \\
$\Delta m_{\tilde{l}_R}$ & 4.8 & 0.05 (LC input)  \\
$\Delta m_{\tilde{l}_L}$ & 5.0 & 0.2 (LC input) \\
$\Delta m_{\tilde{\tau}_1}$ & 5-8 & 0.3 (LC input) \\
$\Delta m_{\tilde{q}_L}$ & 8.7  & 4.9  \\
$\Delta m_{\tilde{q}_R}$ & 7-12 & 5-11  \\
$\Delta m_{\tilde{b}_1}$ & 7.5 & 5.7  \\
$\Delta m_{\tilde{b}_2}$ & 7.9 & 6.2  \\
$\Delta m_{\tilde{g}}$ & 8.0 & 6.5  \\
\hline
\end{tabular}\\[0.2cm]
{\footnotesize The RMS values of the mass distribution in the case of
the LHC alone, and combined with measurements from the LC.
All numbers in GeV.} 
\end{minipage}\\[0.5cm]

The plot in Fig.2 shows clearly the above-mentioned correlations. An accurate 
determination  of the $m_{\tilde{\chi}_1^0}$, with the precision indicated by 
the two vertical lines in Fig.~2 at the ILC, will certainly reduce the error
in the determination of, for example,  $m_{\tilde{b}_1}$.  The numbers in Table 
show how an accurate input from the ILC on masses of the sparticles
{\it that are accessible to the ILC}, will be able to substantially improve 
the accuracy of the mass determination at the LHC for those whose masses
{\it  are beyond the reach of the ILC}. The jet measurement seems to be the 
limiting factor for the accuracies possible with a combined analysis of the 
LHC  and the ILC data. {\it This is an example where the study has isolated 
a feature of LHC analysis which could be improved upon, so as to increase 
the overall precision of sparticle mass determination at the LHC.}
The sparticle masses so determined can then be used to determine the 
pattern of SUSY breaking~\cite{benetal}. 
\setcounter{figure}{2}
\begin{figure}
\centerline{
      \includegraphics*[width=6cm,height=4.8cm]{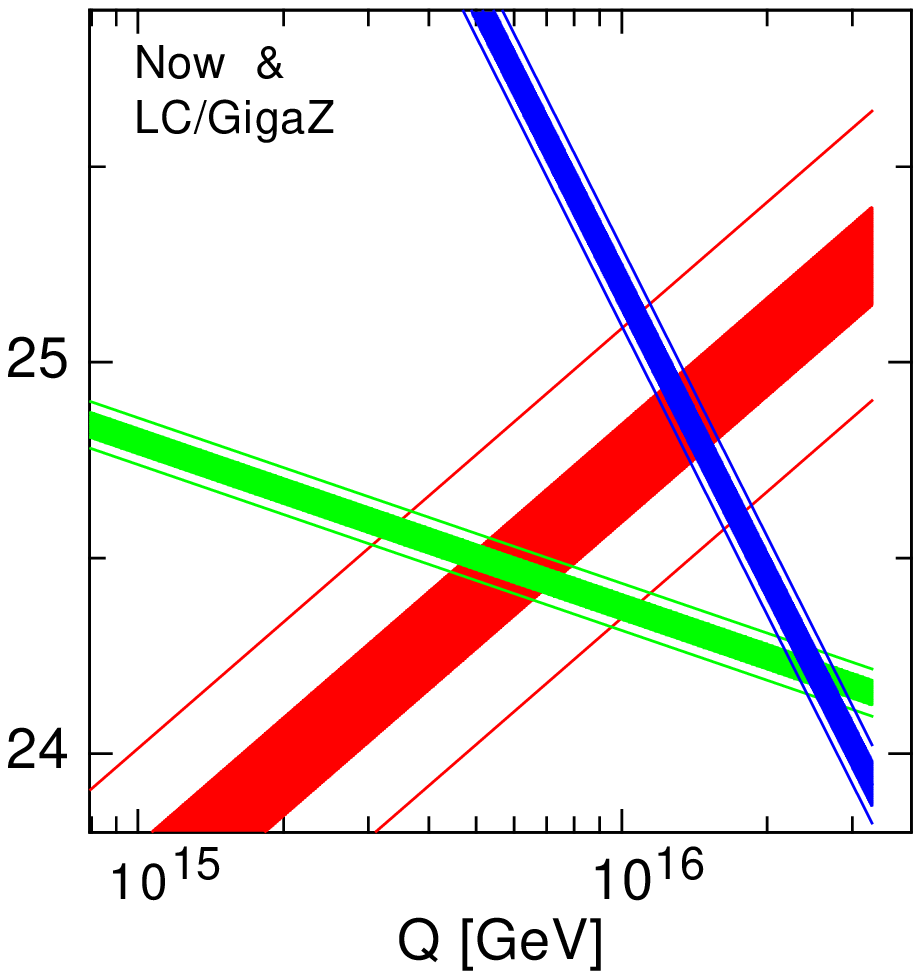}
\hspace{0.3cm}
      \includegraphics*[width=6cm,height=4.5cm]{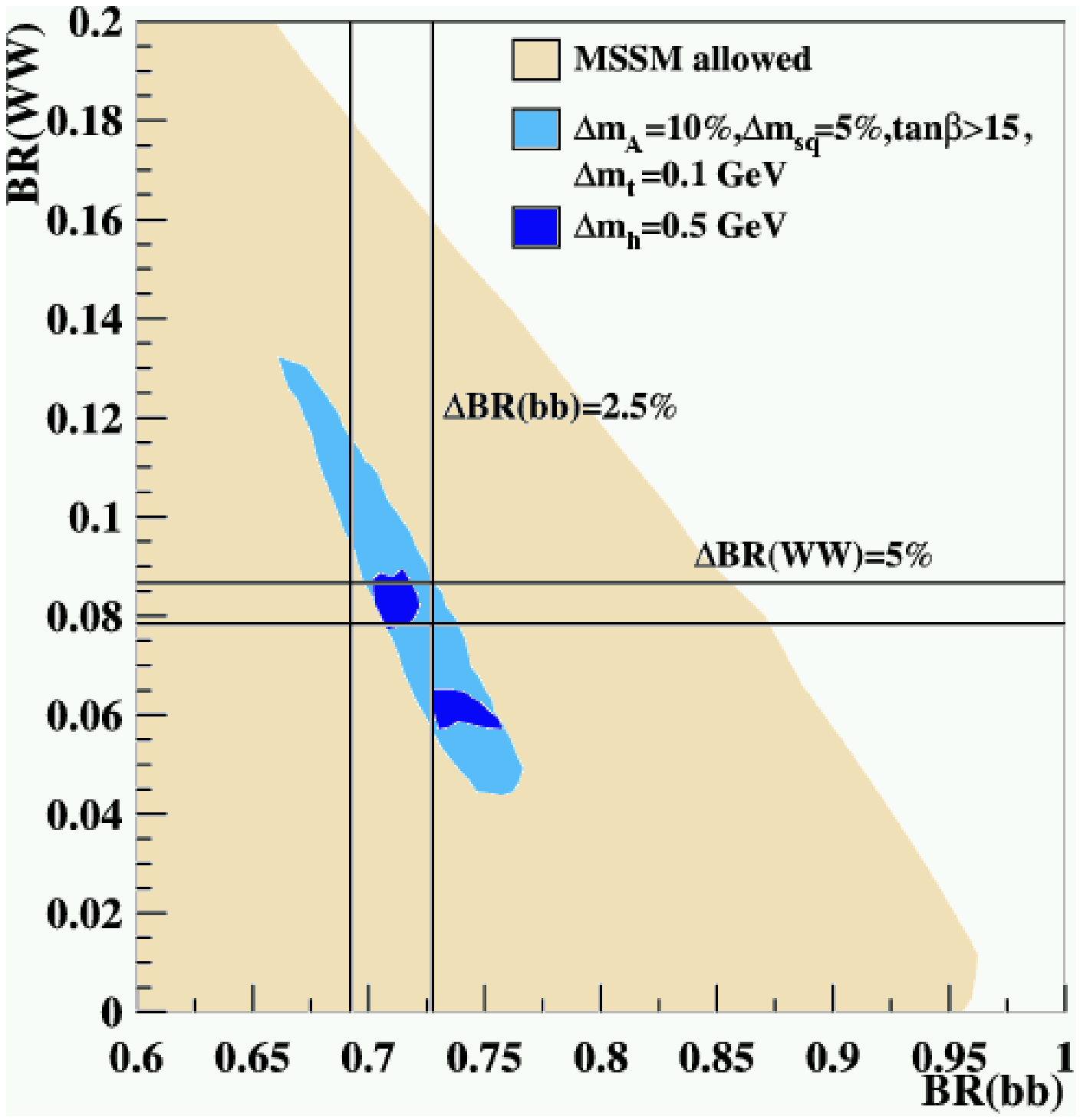}
}
\caption{\footnotesize 
Left panel shows  area around the unification point $M_U$
defined by the meeting point of $\alpha_1$ with $\alpha_2$.
Right hand panel shows the experimental accuracies for the branching 
ratios BR($h \to b \bar b$) and BR($h \to WW^*$) at the LC, indicated by a
vertical and horizontal band, respectively  in comparison with 
the theoretical prediction in the MSSM.}
\end{figure}

The wide error bands in the left panel of Fig. 3 around $M_U$ are based 
on present data, 
and the the expected errors on the spectrum of supersymmetric particles from 
LHC measurements within mSUGRA. The narrow bands demonstrate the improvement 
expected by future GigaZ analyses and the measurement of the complete 
spectrum at ``LHC+LC''.  Thus LHC/ILC synergy can indeed help sharpen our 
knowledge of the High Scale physics beyond the SM. As a matter of fact an 
analysis~\cite{weinmann} for the SPS1a point shows 
that {\it no convergence in the global fits to the MSSM parameters is 
possible without including the ILC/LHC results}.

One very interesting demonstration of the feedback from ILC into LHC studies 
and vice versa was seen in a study of heavier neutralino/chargino sector and
the SUSY parameter determination~\cite{gudi}. They look at a point where the
$\tilde{\chi}_4^0$ can be produced only at the LHC. The heavier states are 
notoriously difficult to study at the LHC.  The measurements of the
$\tilde{\chi}_i^\pm, \tilde{\chi}_i^0, i=1,2$ at the ILC can determine the SUSY
parameters in a {\it model independent} way and  the $m_{\tilde{\chi}_4^0}$ is
then {\it predicted. The ILC can thus tell the  LHC  where to look.}
Armed with this knowledge, the LHC analysis to search for $\tilde{\chi}_4^0$
can be tuned better. The study in the context of  ATLAS detector shows
that the error in  $m_{\tilde{\chi}_4^0}$  determination 
at the LHC can  go down from 5 GeV to 2.5 GeV, for a $m_{\tilde{\chi}_4^0} 
= 378.3$ GeV,\cite{gudi} if such information is available from the ILC to the
LHC analysis. Further the value of $m_{\tilde{\chi}_4^0}$ 
determined at the LHC can then be fed back into the ILC analysis thus
increasing the accuracy of the SUSY parameter determination there.  This 
is need the case of information from both the colliders feeding back into
the study at the other. In another study~\cite{mihoko} a strategy for 
using LHC/ILC together for a determination of the mixing parameters for the 
third generation of squarks at the LHC has been outlined. 

\noindent \underline{SUSY-Higgs at the LHC/ILC}

\vspace{0.1cm}

\noindent In turn the last mentioned  information is very significant for 
a precision prediction of the mass of the 'light' higgs ,$h$,  in SUSY and its 
branching ratios.  A study~\cite{desch2} shows how the combined knowledge can 
improve the significance of testing the consistency of the MSSM from accurate 
measurements of the B.R. of  $h$. This is illustrated in the right 
panel of Fig. 3.  The medium shaded (light blue) region indicates the 
range of predictions in the MSSM being compatible with the assumed 
experimental information from the LHC and the ILC.

This is a case where the accurate $m_t$ determination and the precision 
measurements of the B.R. of the $h$ from the ILC along with the information 
available from the strongly interacting sparticle sector (out of the reach of 
ILC) together can give information on the SUSY parameters such as the trilinear
coupling $A_t$. This in turn can give pointers to the SUSY phenomenology at 
the LHC.  This again is an excellent example of the synergy.

In case of non-universal gaugino masses, SUSY can also make the lightest 
higgs $h$ 'invisible' due to decays into neutralinos. In this case the 
$\tilde{\chi}_1^0$ necessarily has a substantial higgsino component, 
affecting the sparticle search at the LHC. Detection of such a $h$
at the LHC is difficult, if not impossible. 
\begin{figure}
\centerline{
      \includegraphics*[scale=0.30]{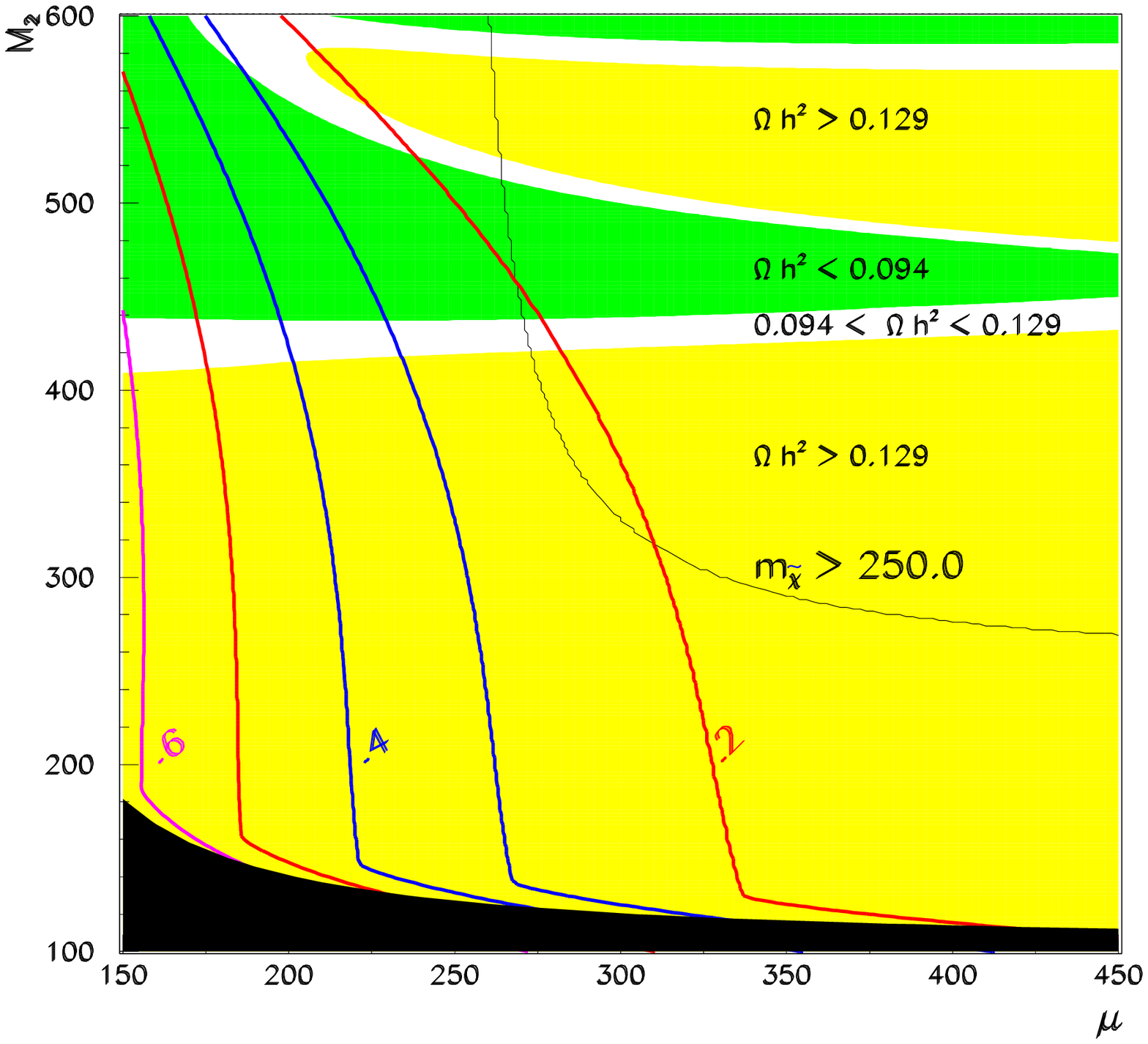}
      \includegraphics*[scale=0.65]{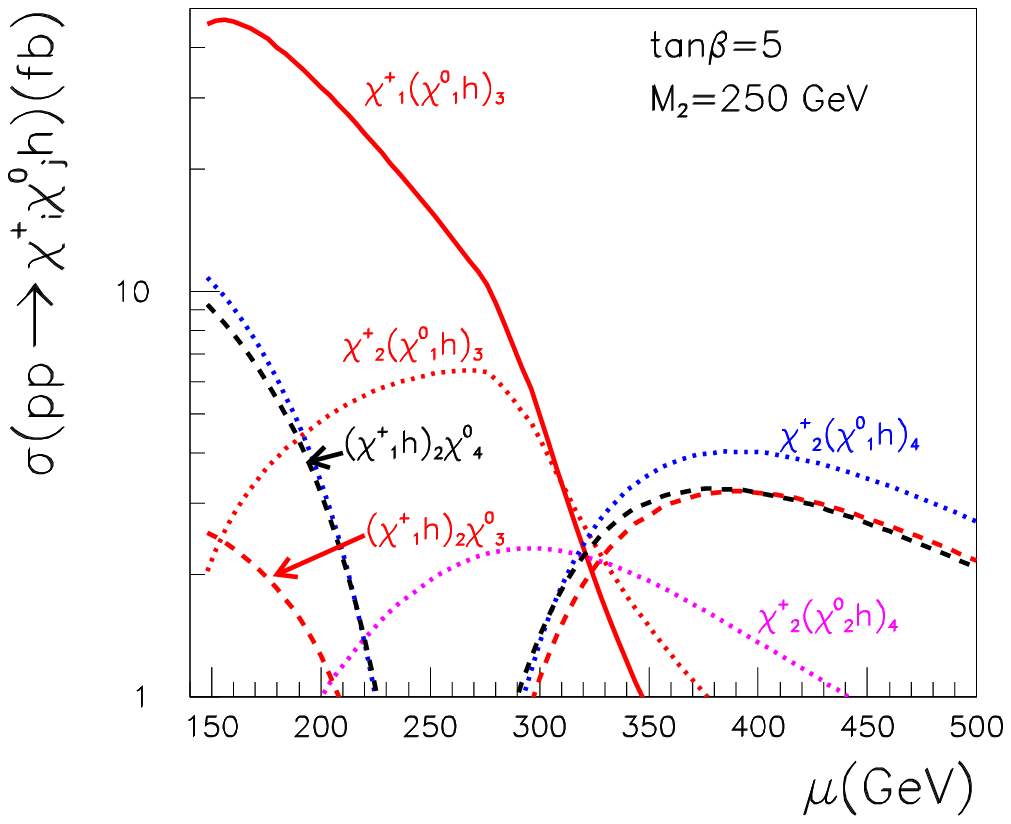}
}
\caption{\footnotesize
The left panel shows contours of 'invisible' branching ratio of
the $h$i, for a particular choice of the non-universality of the
Gaugino masses, along with the DM and LEP constraints.
Higgs yield through charginos and neutralinos decays as a function of 
$\mu$.  The subscript for the parentheses $(\;\;)_j$ indicates the 
parent neutralino or chargino\/.}
\end{figure}
As shown in the left panel of Fig.4
cosmology constraints disfavour a large region in the parameter space where
this can happen~\cite{fawzi}. Even then, substantial portions of the 
parameter space 
where this may happen are still allowed. Needless to say that an ILC can detect
a $h$ with 'invisible' decay products quite easily. If the 'invisibility' is 
due to SUSY, one expects enhanced production of $h$ in $\tilde{\chi}_i^0$ 
decays, as shown in the right panel of Fig. 4. Thus this is a case where ILC 
input will play a useful role in pinning down the SUSY scenario.

\section*{Contact interactions, new gauge theories:}
Due to the developments in the Little Higgs Models there has been new 
impetus to look at theories with an extended gauge sector. Equally important 
are the new ideas of the Extra Dimensional models which predict Kaluza Klein
(KK) excitation of the gauge bosons. A result of the investigation of the
possibility of discriminating between these different models which predict 
extra gauge bosons and the role of the LHC/ILC synergy in this 
is present in the report~\cite{Bourilkov}. Again, it demonstrates ample 
scope for the LHC/ILC synergy. For example, LHC can see new resonances, 
at a mass that is not accessible to the ILC whereas the ILC using the 
LHC pointers can measure couplings through a simple study of multi-fermion 
final states. One can then use the precision measurements at a GigaZ to 
distinguish between these different models. 
\begin{figure}
\centerline{
      \includegraphics*[width=6cm,height=4cm]{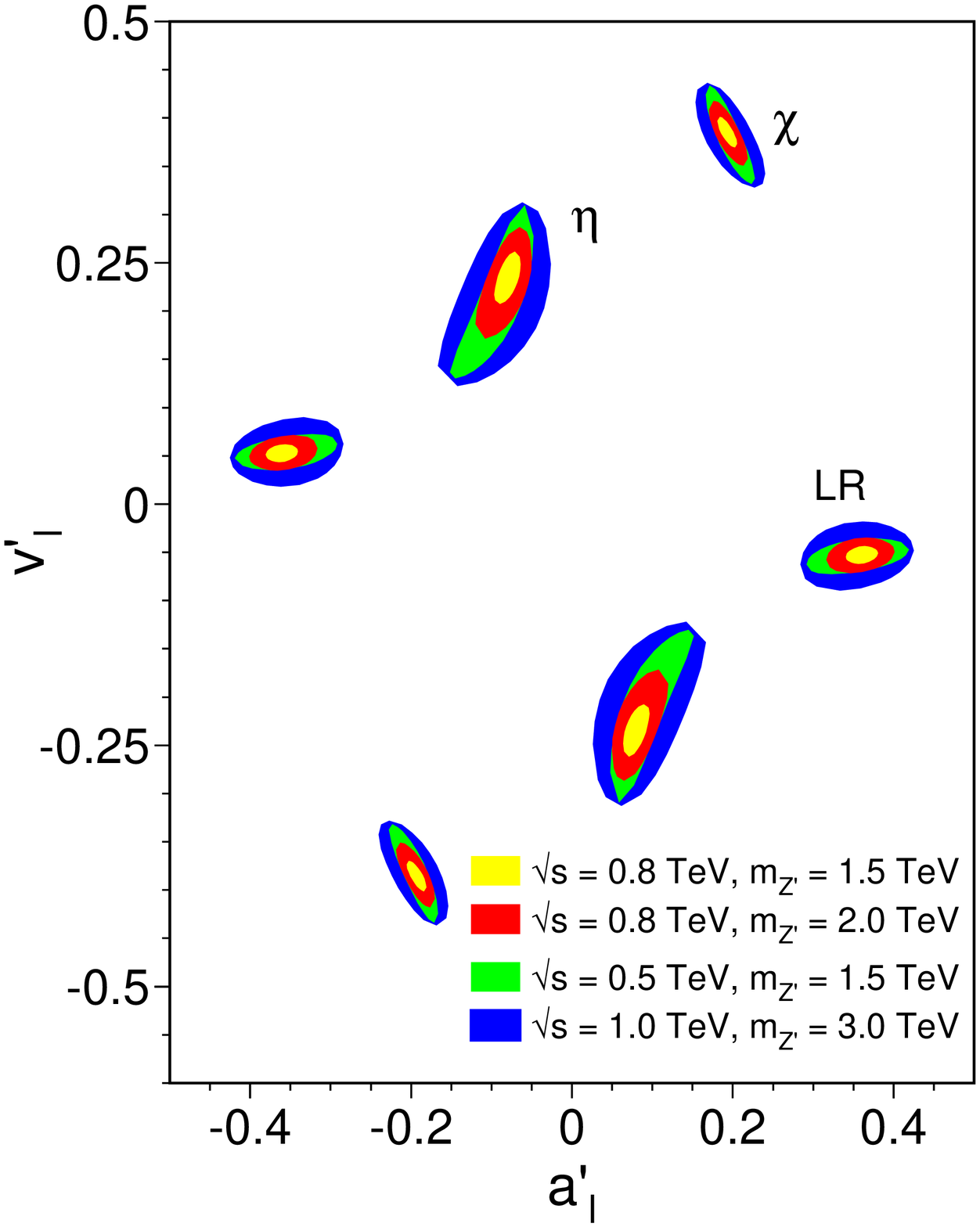}
\hspace{0.5cm}
      \includegraphics*[width=6cm,height=4cm]{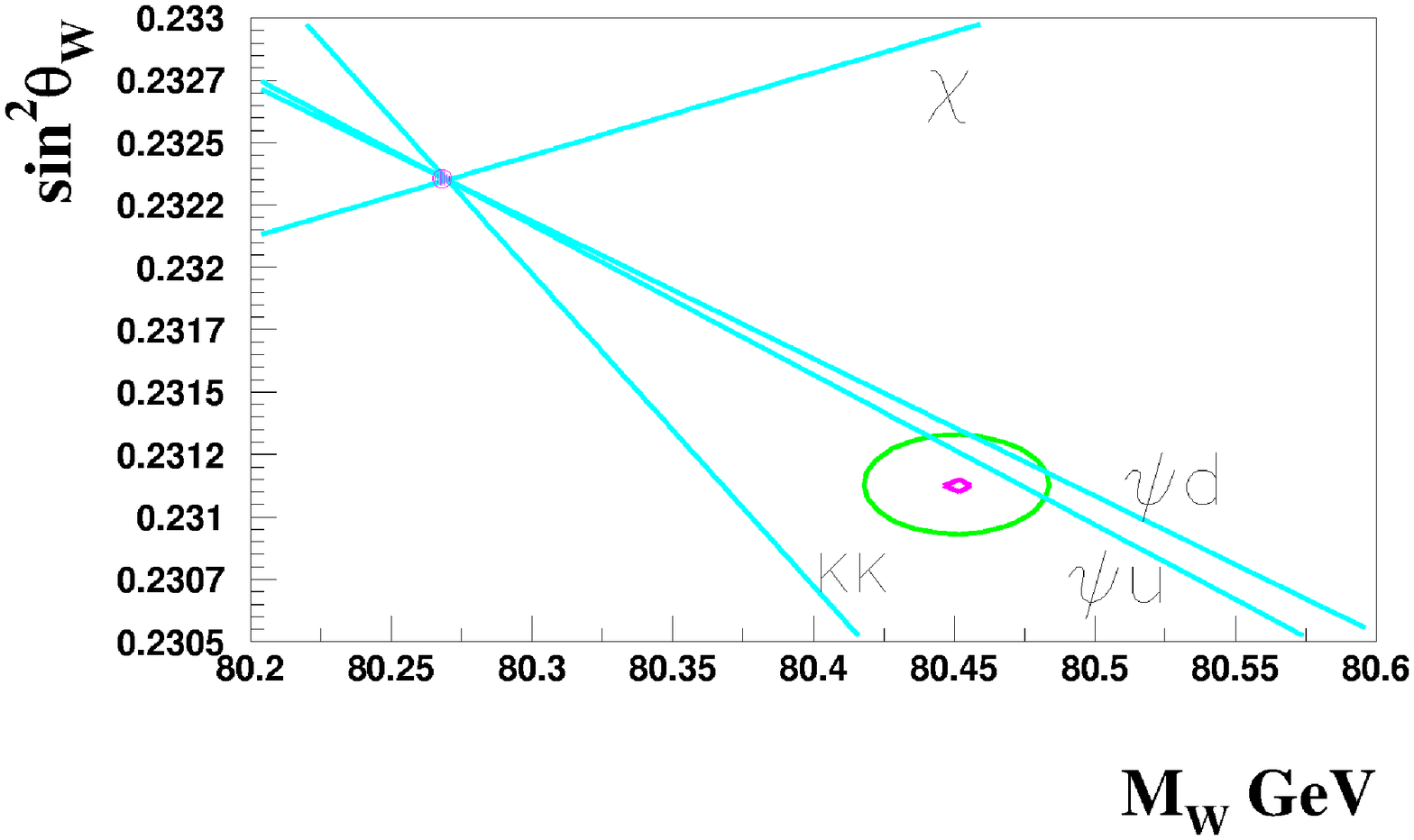}
}
\caption{\footnotesize Left panel indicates The 95\% C.L. contours on 
leptonic Z$'$ couplings, assuming the mass of the Z$'$ is measured at LHC.
SM prediction, LEP/SLD  and GigaZ expected precision compared with Z$'$ 
models. The ellipse in the lower part of the Fig. corresponds to the current 
experimental accuracy where as the inner one indicate reach at the GigaZ. The 
left, upper ellipse indicates the SM prediction.}
\end{figure}

The left panel of the Fig.5 indicate the 
ability of an ILC to distinguish between the different models by a  measurement
of the couplings of a new $Z'$ ($m_{Z'} < 4.5 $ TeV) for which the LHC has 
a reach, whereas the right panel indicates the ability to do so through
the EW precision measurements of $\sin^2\theta_W, m_W$. The much narrower 
inner circle indicates the reach of a GigaZ.

\section*{Conclusions}
Only a very small sample of the LHC/ILC study group document was presented 
here. The document contains many more examples, in 1)EW physics, 2)QCD and Top
Physics, 3) Studies of the Higgs Potential 4) CP studies in the Higgs sector, 
5)Extra dimensional models, 7) Radion-Higgs separation, 8)Little Higgs studies,
9)NonMinimal Supersymmetric Standard Model (NMSSM) etc., where the LHC/ILC
synergy has been demonstrated in a {\it quantitative} fashion.  
New points for further studies have been also been identified. In some cases, 
fully simulated LHC and ILC events have been used for analysis. Many are 
still studies by phenomenologists which need experimental simulations. This 
document is but just a beginning having scratched only the surface. Certainly
the more and new studies need to be performed. However, the LHC/ILC study
document has indeed shown that it is hard to believe that after
the ILC turns on, no new questions will be asked of the LHC. All this points
towards the need of having some {\it overlap} in the running life of the
two machines the LHC and the ILC.

\vspace{-0.2cm}
\section{Acknowledgements}
\vspace{-.3cm}
It is  a pleasure to thank H. Videau for organising the 'colloq'. Special
thanks go to G. Weiglein for organisation of the activities of the ILC/LHC
Study Group and the Study Group Document. I would like to acknowledge the 
support of Department of Science and Technology, India, under project 
no. SP/S2/K-01/2000-II.

\end{document}